\documentclass[%
 nofootinbib,
 amsmath,amssymb,
 aps,
 prd,
twocolumn, superscriptaddress
]{revtex4-1}

\usepackage{float}
\usepackage[normalem]{ulem}

\usepackage{graphicx}
\usepackage{dcolumn}
\usepackage{bm}
\usepackage{hyperref}

\usepackage{xcolor}

\usepackage{color}
\definecolor{DarkGreen}{rgb}{.2, .6, .2}

\begin{document}

\preprint{APS/123-QED} 

\title{ Charged quark stars in metric $f(R)$ gravity }

\author{Juan M. Z. Pretel}
 \email{juanzarate@cbpf.br}
 \affiliation{
 Centro Brasileiro de Pesquisas F{\'i}sicas, Rua Dr. Xavier Sigaud, 150 URCA, Rio de Janeiro CEP 22290-180, RJ, Brazil
}

\author{Jos\'e D. V. Arba\~nil}
\email{jose.arbanil@upn.pe}
\affiliation{Departamento de Ciencias, Universidad Privada del Norte, Avenida el Sol 461 San Juan de Lurigancho, 15434 Lima,  Peru}
\affiliation{Facultad de Ciencias F\'isicas, Universidad Nacional Mayor de San Marcos, Avenida Venezuela s/n Cercado de Lima, 15081 Lima,  Peru
}

\author{Sergio B. Duarte}
 \email{sbd@cbpf.br}
 \affiliation{
 Centro Brasileiro de Pesquisas F{\'i}sicas, Rua Dr. Xavier Sigaud, 150 URCA, Rio de Janeiro CEP 22290-180, RJ, Brazil
}

\author{Sergio E. Jor\'as}
 \email{joras@if.ufrj.br}
 \affiliation{
 Instituto de F\'\i sica, Universidade Federal do Rio de Janeiro,\\
 CEP 21941-972 Rio de Janeiro, RJ, Brazil 
}

\author{Ribamar R. R. Reis}
 \email{ribamar@if.ufrj.br}
\affiliation{
 Instituto de F\'\i sica, Universidade Federal do Rio de Janeiro,\\
 CEP 21941-972 Rio de Janeiro, RJ, Brazil 
}
\affiliation{Universidade Federal do Rio de Janeiro, Observat\'orio do Valongo, 
 \\CEP 20080-090 Rio de Janeiro, RJ, Brazil}

\date{\today}

\begin{abstract}
We provide the modified TOV equations for the hydrostatic equilibrium of charged compact stars within the metric $f(R)$ gravitational background. We adopt the MIT bag model EoS for the dense matter and assume a charge distribution where the electric charge density $\rho_{\rm ch}$ is proportional to the standard energy density $\rho$. Using the Starobinsky model, we explore the role of the $\alpha R^2$ term, where $\alpha$ is a free constant and $R$ the Ricci scalar, on the global properties of charged stars such as radius, mass and total charge. We present the dependence of the structure of the star for several values of $\alpha$ and for different values of the constant parameter $\beta\equiv \rho_{\rm ch}/\rho$. Remarkably, we find that the radius decreases with respect to its GR value for low central densities, while the opposite occurs in the high-central-density region. The mass measured at the surface always decreases and the maximum-total charge undergoes a substantial increase as the parameter $\alpha$ increases. We also illustrate the variations of the asymptotic mass as a consequence of the electric charge and the extra quadratic term.

\end{abstract}

\maketitle


\section{Introduction}

It has been argued that strange stars might contain a layer of electrons (called the electrosphere) and a quark region, which are bound together by an extremely strong electric field \cite{Alcock1986, Usov2004, Usov2005}. In that regard, the inclusion of electromagnetic field in the energy-momentum tensor is of crucial importance in the construction of compact stars. Within the framework of General Relativity (GR from now on), the effect of electric charge on compact stars using a polytropic equation of state (EoS) was investigated in Refs.~\cite{Ray2003, Arbanil2013}, where the authors assumed that the charge distribution is proportional to the energy density (see also Ref.~\cite{Arbanil2014} for the study of relativistic charged spheres made of an incompressible fluid). Weber and collaborators \cite{Negreiros2009, Weber2010} showed that electric charge distributions located near the surface of strange quark stars increase the stellar mass by up to $15\%$ and the radius by up to $5\%$, depending on the amount of electric charge carried by the star. Based on a power-law charge distribution \cite{Felice1995, Felice1999}, the stability analysis against radial pulsations of charged quark stars was carried out in Refs.~\cite{Arbanil2015, Zhang2021}, where it was considered that the charge is proportional to spatial volume. Additional results on the impact of electric charge on the global properties of compact stars can be found in Refs.~\cite{Thirukkanesh2017, Carvalho2018, Kumar2019, Panotopoulos2019, Lazzari2020, Ivanov2021, Jasim2021, PANOTOPOULOS2021, Delgado2022}.

Electrically charged compact stars were also investigated in some modified theories of gravity. As a matter of fact, charged compact stars with polytropic EoS in the Starobinsky model were constructed in Ref.~\cite{Mansour2018}. Rocha \textit{et al.}~\cite{Rocha2020} explored white dwarfs composed of a charged perfect fluid in $f(R,T)$ gravity. See also Ref.~\cite{Rej2021} for the study of charged stars using the metric potentials proposed by Tolman–Kuchowicz and Ref.~\cite{Deb2019} for charged strange stars in the same theory. In addition, the literature provides studies on charged stars in other theories of gravity \cite{Ilyas2018, Shahzad2019, Bhatti2021, Pretel2022, Panotopoulos2022, Salako2022}.


The study of modified theories of gravity aims, usually, to replace the Cosmological Constant as the responsible for the current accelerated expansion of the Universe. Nevertheless, it can also be used to investigate --- and, therefore, to constrain --- alternative theories that may come from any high-energy corrections to GR (e.g., string theory). In this paper we focus on an extensively used modified theory, namely, the Starobinsky model ($f(R) = R + \alpha R^2$, in the metric approach \cite{Starobinsky1980}) and add two extra ingredients: a quark star with electric charge. The goal is twofold: we wish to find out if modifications of both GR and of the net charge can help to (de)stabilize such a star. Besides, the addition of electric charge can also be seen as a first look into extra degrees of freedom, not covered by $f(R)$ theories.

In $f(R)$ gravity, both the metric and the Ricci scalar are dynamical fields governed by differential equations \cite{Olmo2007, SotiriouFaraoni, Felice}. Namely, the scalar curvature plays a nontrivial role in the determination of the metric itself. In fact, the algebraic equation $R= -8\pi T$ (obtained in Einstein gravity) is no longer valid. In particular, in the exterior spacetime of a compact star, where $T= 0$, the scalar curvature $R$ may not vanish  and hence the Reissner-Nordstr{\"o}m metric is inadequate. See Refs.~\cite{Yazadjiev2014, Yazadjiev2015, Astashenok2015, Astashenok2017, Sbisa2020, Astashenok2020, Astashenok2021, Jimenez2021} for the uncharged case of compact stars within the context of $f(R)$ gravity, where it has been shown that the scalar curvature does not vanish in the outer region of a compact star. However, the authors of Ref.~\cite{Mansour2018} assumed that the interior solution is smoothly connected to the exterior Reissner-Nordstr{\"o}m metric by considering $R(r_{\rm sur})= 0$, where $r_{\rm sur}$ is the radius of the star, which we believe to be a too strong constraint. To the best of our knowledge, charged compact stars have not yet been explored in metric $f(R)$ gravity by satisfying the second-order differential equation for $R$ and with suitable boundary conditions, so the purpose of the present work is to fill this gap by adopting a non-perturbative approach for the field equations.

The present study is organized as follows: In Sec. \ref{Sec2} we briefly review $f(R)$ gravity theories and present the field equations for a static spherically symmetric system in the presence of electric charge. In the same section we provide an explicit expression for the gravitational mass and derive the modified Tolman-Oppenheimer-Volkoff (TOV) equations describing charged fluid spheres. Section \ref{Sec3} presents a well-known EoS to describe quark stars as well as a charge-distribution model. In Sec. \ref{Sec4} we discuss our numerical results, and finally, our conclusions are presented in Sec. \ref{Sec5}. In this work we will use the sign convention $(-,+,+,+)$ and a geometric units system --- our results, however, will be given in physical units for comparison purposes.


\section{Stellar structure equations}\label{Sec2}

We study the hydrostatic equilibrium structure of electrically charged compact stars in $f(R)$ theories of gravity within the metric formalism. In this section we briefly summarize the gravitational background by taking into account the Maxwell contributions. In particular, we will focus on the case where the magnetic field vanishes, but there exists a static electric field due to an electric charge source in the stellar fluid.

\subsection{Field equations}

In the presence of electromagnetic field, the most general action for $f(R)$ modified gravity is defined by 
\begin{equation}\label{action}
    S = \int d^4x\sqrt{-g} \left[ \frac{1}{16\pi}f(R) + \mathcal{L}_m + \mathcal{L}_e \right] ,
\end{equation}
where $g$ is the determinant of the metric, $\mathcal{L}_m$ stands for the Lagrangian density for the standard matter distribution and $\mathcal{L}_e$ denotes the Lagrangian for the
electromagnetic field. By varying the action (\ref{action}) with respect to the metric we obtain the field equations:
\begin{equation}\label{FielEq}
f_R R_{\mu\nu} - \dfrac{1}{2}g_{\mu\nu}f - \nabla_\mu\nabla_\nu f_R + g_{\mu\nu}\square f_R = 8\pi T_{\mu\nu} ,
\end{equation}
where $f_R \equiv df(R)/dR$ and $\square \equiv \nabla_\mu\nabla^\mu$ is the d'Alembert operator with $\nabla_\mu$ representing the covariant derivative. The field equations in $f(R)$ theories of gravity are fourth-order differential equations in the metric components, and the conventional Einstein-Maxwell equation is retrieved when $f(R)= R$. Notice also that the metric is generated by the matter-energy distribution and by the terms related to the scalar curvature.

The total energy-momentum tensor in Eq.~(\ref{FielEq}) is the sum of two terms, namely $T_{\mu\nu} = \mathcal{M}_{\mu\nu} + \mathcal{E}_{\mu\nu}$, where $\mathcal{M}_{\mu\nu}$ corresponds to a perfect fluid and whose energy-momentum tensor in the comoving frame can be written as
\begin{equation}\label{EMTensor1}
    \mathcal{M}_{\mu\nu} = (\rho + p)u_\mu u_\nu + pg_{\mu\nu}, 
\end{equation}
with $u^\mu$ being the four-velocity of the fluid, $\rho$ the energy density and $p$ is the pressure. 

In addition, the electromagnetic energy-momentum tensor $\mathcal{E}_{\mu\nu}$ has the following form 
\begin{equation}\label{EMTensor2}
    \mathcal{E}_{\mu\nu} = \frac{1}{4\pi} \left[ F_{\mu\lambda}F_\nu^{\ \lambda} - \frac{1}{4}g_{\mu\nu}F_{\lambda\sigma}F^{\lambda\sigma} \right] ,
\end{equation}
where $F_{\mu\nu} = \partial_\mu A_\nu - \partial_\nu A_\mu$ is the antisymmetric electromagnetic field strength tensor and $A_\mu$ is the electromagnetic four-potential. It should be noted that the energy-momentum tensor (\ref{EMTensor2}) has zero trace, \begin{equation}\label{5}
    \mathcal{E} = \mathcal{E}_\mu^{\ \mu} = \frac{1}{4\pi}\left[ F_{\mu\lambda}F^{\mu\lambda} - \frac{1}{4}\delta_\mu^{\ \mu} F_{\lambda\sigma}F^{\lambda\sigma} \right] = 0, 
\end{equation}
so that $T = g_{\mu\nu}T^{\mu\nu} = \mathcal{M} = -\rho+ 3p$. Consequently, the trace of Eq.~(\ref{FielEq}) leads to a second-order differential equation for the Ricci scalar
\begin{equation}\label{TraceEq}
    3\square f_R(R) + Rf_R(R) - 2f(R) = 8\pi \mathcal{M} .
\end{equation}

Unlike the pure GR case, where the scalar curvature is governed by an algebraic equation (namely, $R= -8\pi \mathcal{M}$ and hence $R= 0$ in the absence of matter), in $f(R)$ gravity both $g_{\mu\nu}$ and $R$ are dynamical fields determined by the differential equation (\ref{TraceEq}). In other words, non-linear Lagrangian functions in $R$ allow for a non-zero scalar curvature even in the exterior region of a compact star where $\mathcal{M}= 0$.

The electromagnetic strength tensor must satisfy the Maxwell equations:
\begin{align}
    &\frac{1}{\sqrt{-g}} \partial_\mu\left( \sqrt{-g}F^{\mu\nu} \right) = -4\pi j^\nu ,  \label{MEq1}  \\
    &\nabla_\sigma F_{\mu\nu} + \nabla_\mu F_{\nu\sigma} + \nabla_\nu F_{\sigma\mu} = 0 ,  \label{MEq2}
\end{align}
where $j^\mu = \rho_{\rm ch}u^\mu$ is the four-current density and $\rho_{\rm ch}$ is the electric charge density.

Let us consider a static spherically symmetric system whose spacetime is described by the following line element
\begin{equation}\label{MetricEq}
    ds^2 = -e^{2\psi}dt^2 + e^{2\lambda}dr^2 + r^2(d\theta^2 + \sin^2\theta d\phi^2) ,
\end{equation}
where $x^\mu = (t,r,\theta,\phi)$ are the components of the four-position vector, and the metric functions $\psi$ and $\lambda$ depend only on the radial coordinate $r$. This implies that $\sqrt{-g}= e^{\psi+ \lambda}r^2\sin\theta$ and $u^\mu = (e^{-\psi}, 0, 0, 0)$. Under the assumption that the electromagnetic field is generated solely due to the electric charge, the only non-zero component of the strength tensor is $F^{01} = -F^{10}$. Therefore, the Maxwell equation (\ref{MEq1}) reduces to
\begin{equation}
    F^{01} = \frac{q(r)}{r^2}e^{-\psi - \lambda} , 
\end{equation}
where we have defined $q(r)$ as a charge function inside a sphere of radius $r$, given by 
\begin{equation}\label{ChargeEq}
    q(r) = 4\pi\int _0^r \bar{r}^2\rho_{\rm ch}(\bar{r})e^{\lambda(\bar{r})}d\bar{r} .
\end{equation}
From now on, we will not explicitly indicate the $r$ dependence on $q(r)$ (nor\ on pressure $p$ and densities $\rho$ and $\rho_{\rm ch}$) to avoid cluttering the equations.

From Eqs.~(\ref{EMTensor2}) and (\ref{MEq2}), we find that $\nabla_{\mu}\mathcal{E}^{\mu\nu} = j_{\lambda} F^{\lambda\nu}$. Subsequently, the four-divergence of the tensor $T^{\mu\nu}$ provides the conservation law of energy and momentum as in the standard GR theory, namely
\begin{equation}\label{ConservationEq}
    \nabla_\mu T_1^{\ \mu} = p' + (\rho + p)\psi' - \frac{q}{r^2}\rho_{\rm ch}e^\lambda = 0 ,
\end{equation}
where the prime represents differentiation with respect to the radial coordinate. Besides, the dynamical equation for the Ricci scalar (\ref{TraceEq}) becomes 
\begin{align}\label{RicciScalarEq}
    \frac{3}{e^{2\lambda}}\left[ \left( \frac{2}{r} + \psi' - \lambda' \right)f_R' + f_R^{''} \right] &+ Rf_R - 2f  \nonumber  \\
    =&\ 8\pi(-\rho + 3p) .
\end{align}

In view of the energy-momentum tensors (\ref{EMTensor1}) and (\ref{EMTensor2}), together with the metric (\ref{MetricEq}), the non-zero components of the field equations (\ref{FielEq}) are given by
\begin{widetext}
\begin{eqnarray}
    &&-\frac{f_R}{r^2} + \frac{f_R}{r^2}\frac{d}{dr}\left( re^{-2\lambda} \right) + \frac{1}{2}(Rf_R - f) + \frac{1}{e^{2\lambda}}\left[ \left( \frac{2}{r} - \lambda' \right)f_R' + f_R^{''} \right] = -8\pi \left( \rho + \frac{q^2}{8\pi r^4} \right), \label{FielEq1}
\\    
&&-\frac{f_R}{r^2} + \frac{f_R}{e^{2\lambda}}\left( \frac{2\psi'}{r} + \frac{1}{r^2} \right) + \frac{1}{2}(Rf_R- f) + \frac{1}{e^{2\lambda}}\left( \frac{2}{r} + \psi' \right)f_R' = 8\pi \left( p - \frac{q^2}{8\pi r^4} \right), \label{FielEq2}
\\    
&&\frac{f_R}{r^2}\left[ \frac{1}{e^{2\lambda}}(r\lambda' - r\psi' -1) +1 \right] - \frac{1}{2}f + \frac{1}{e^{2\lambda}}\left[ \left( \frac{1}{r} + \psi' - \lambda' \right)f_R' + f_R^{''} \right] = 8\pi \left( p + \frac{q^2}{8\pi r^4} \right), \label{FielEq3}
\end{eqnarray}
\end{widetext}
which are reduced to the pure general relativistic case \cite{Ray2003} when $f(R)= R$.

\subsection{Mass function}

In order to derive an explicit expression for the gravitational mass of a charged compact star in $f(R)$ gravity, we will follow a procedure analogous to that carried out in Einstein gravity \cite{Arbanil2013}. Namely, from the 00-component of the field equations (\ref{FielEq1}) we obtain 
\begin{align}\label{Mf1}
    \frac{d}{dr}\left( re^{-2\lambda} \right) =&\ 1- 8\pi r^2\rho - \frac{q^2}{r^2} - \left\lbrace \frac{r^2}{2}(Rf_R- f)   \right.  \nonumber  \\
    &\left. + (1- f_R)\frac{d}{dr}\left[ r(1- e^{-2\lambda}) \right]  \right.  \nonumber  \\
    &\left. + \frac{r^2}{e^{2\lambda}}\left[ \left( \frac{2}{r}- \lambda' \right)f'_R+ f''_{R} \right]  \right\rbrace ,
\end{align}
whose integration leads to the following  equation
\begin{equation}\label{Mf2}
    e^{-2\lambda} = 1 - \frac{2m}{r} + \frac{q^2}{r^2} ,
\end{equation}
where $m(r)$ plays the role of mass function and determines the gravitational mass enclosed within the radial coordinate $r$. In the study of relativistic charged spheres, Bekenstein \cite{Bekenstein1971} introduced the expression (\ref{Mf2}) so that the spacetime outside of a non-rotating charged source is described by the Reissner-Nordström metric in the Einstein-Maxwell theory.

Nonetheless, in the context of $f(R)$ modified gravity, the standard Reissner-Nordström vacuum solution is invalid due to the presence of the extra term inside the curly braces in Eq.~(\ref{Mf1}). 

As a consequence, bearing in mind that $f_R' = R'f_{RR}$ and $f_R^{''} = R''f_{RR} + R'^2f_{RRR}$, the gravitational mass --- i.e., one measured by a distant observer\footnote{See, however, other distinct mass definitions in Ref.~\cite{Sbisa2020}} --- can be written as
 \begin{align}\label{Mf3}
     m(\bar{r}) =&\ 4\pi\int_{0}^{\bar{r}} \rho r^2 dr + \int_{0}^{\bar{r}}\frac{qq'}{r}dr + \frac{1}{2}\int_{0}^{\bar{r}}\left\lbrace  \frac{1}{2}(Rf_R- f) \right.  \nonumber  \\
     &\left. + \frac{1}{e^{2\lambda}}\left[ \left( \frac{2}{r}- \lambda' \right)R'f_{RR} + R''f_{RR}+ R'^2f_{RRR} \right] \right.  \nonumber  \\
     &\left. + \frac{(1- f_R)}{r^2}\frac{d}{dr}\left[ r(1- e^{-2\lambda}) \right] \right\rbrace r^2dr . 
 \end{align}
Equation (\ref{Mf3}) means that the mass function is generated by the total contribution from the standard energy density $\rho$, the electric charge $q$ and also from the high-order terms related to the scalar curvature $R$ --- the so-called (effective) curvature fluid. Below we will show a graphical analysis for the energy density associated with this fluid, check Fig.~\ref{figure2}. The uncharged case is retrieved when $q= 0$, as given in Ref.~\cite{PretelSD2022}. We further remark that when $f(R) = R$, the third integral on the right-hand side of the equation above vanishes and hence we recover the typical expression obtained in the pure GR scenario \cite{Arbanil2013, Felice1995, Arbanil2014, Lazzari2020, Ivanov2021, PANOTOPOULOS2021, Zhang2021}. However, notice that even in the exterior region of an electrically charged compact star, where $\rho = 0$, the above expression yields an extra mass contribution due to the non-trivial behaviour of the Ricci scalar.


\subsection{TOV equations and boundary conditions }

The main purpose of the present work is to describe charged stars in $f(R)$ gravity under a non-perturbative treatment of the field equations. In that regard, we need to derive the modified version of the TOV equations in such gravitational context by taking into account the electric charge distribution. Therefore, after suitably combining equations (\ref{ChargeEq})-(\ref{FielEq2}), we get the following differential equations 
\begin{widetext}
\begin{align}
    \frac{d\psi}{dr} &= \frac{1}{2r(2f_R+ rR'f_{RR})}\left\lbrace r^2e^{2\lambda}\left[16\pi \left( p - \frac{q^2}{8\pi r^4} \right) + f - Rf_R\right] + 2f_R\left( e^{2\lambda}- 1 \right) - 4rR'f_{RR} \right\rbrace ,  \label{TOV1}  \\
    \frac{d\lambda}{dr} &= \frac{1}{2r(2f_R+ rR'f_{RR})}\left\lbrace 2f_R\left(1- e^{2\lambda}\right) + \frac{r^2e^{2\lambda}}{3}\left[ 16\pi\left(2\rho + 3p + \frac{3q^2}{8\pi r^4} \right) + Rf_R +f \right] \right.  \nonumber  \\
    &\hspace{2cm} \left. +\frac{rR'f_{RR}}{f_R}\left[ 2f_R\left(1- e^{2\lambda}\right) + \frac{r^2e^{2\lambda}}{3}\left( 16\pi\rho + \frac{6q^2}{r^4} + 2Rf_R - f \right) + 2rR'f_{RR} \right] \right\rbrace  ,  \label{TOV2}   \\
    \frac{d^2R}{dr^2} &= \frac{1}{3f_{RR}}\bigg\lbrace e^{2\lambda}\left[ 8\pi(-\rho + 3p) + 2f -Rf_R \right] -3R'^2f_{RRR} \bigg\rbrace + \left( \lambda' - \psi'- \frac{2}{r} \right)R' ,  \label{TOV3}  \\
    \frac{dp}{dr} &= -(\rho + p)\psi' + \frac{qq'}{4\pi r^4} ,  \label{TOV4}  \\  
    \frac{dq}{dr} &= 4\pi r^2\rho_{\rm ch}e^\lambda ,  \label{TOV5}
\end{align}
\end{widetext}
where the specific case $\rho_{\rm ch}= 0$ corresponds to TOV equations describing uncharged compact stars in $f(R)$ gravity \cite{Yazadjiev2015}. The system of equations (\ref{TOV1})-(\ref{TOV5}) corresponds to four first-order and one second-order ordinary differential equations, where there are seven variables (i.e., $\psi$, $\lambda$, $R$, $\rho$, $p$, $q$ and $\rho_{\rm ch}$) to be determined. 

We have to impose suitable boundary conditions at the center of the star and integrate outwards up to the surface (and beyond for the exterior spacetime). As usual, we will define the stellar surface at the radial coordinate where the pressure vanishes, i.e., when $p(r= r_{\rm sur})= 0$. Adopting a barotropic EoS in the form $p= p(\rho)$ and the electric charge density being written in terms of the energy density as in Refs.~\cite{Arbanil2014, Zhang2021}, then only six initial conditions are required to solve the system (\ref{TOV1})-(\ref{TOV5}) inside the star, namely
\begin{align}\label{BC1}
    \rho(0) &= \rho_c ,   &   q(0) &= 0 ,   &   \psi(0) &= \psi_c ,  \nonumber   \\
    \lambda(0) &= 0 ,   &   R(0) &= R_c ,   &   R'(0) &= 0 ,
\end{align}
where $\rho_c$ and $R_c$ are the central values of the energy density and Ricci scalar, respectively. On the other hand, in the outer region of the star ($r> r_{\rm sur}$), where $\rho =p =0$, the system is described by the set of differential equations \eqref{TOV1}, \eqref{TOV2}, \eqref{TOV3}, and \eqref{TOV5}. Therefore, for the exterior problem we integrate the coupled differential equations up to a large enough distance (see section \ref{Sec4}) by setting the junction conditions at the stellar surface 
\begin{align}\label{BC2}
    \psi_{in} (r_{\rm sur}) &= \psi_{out} (r_{\rm sur}) ,   &   \lambda_{in} (r_{\rm sur}) &= \lambda_{out} (r_{\rm sur}) ,  \nonumber  \\
    R_{in} (r_{\rm sur}) &= R_{out} (r_{\rm sur}) ,   &   R'_{in} (r_{\rm sur}) &= R'_{out} (r_{\rm sur}) ,  \nonumber  \\
    q_{in} (r_{\rm sur}) &\equiv Q = q_{out} (r_{\rm sur}) ,
\end{align}
Accordingly, we assume that the electric charge density also vanishes outside the star. Then, the total charge of the compact star is given by $Q\equiv q(r_{\rm sur})$. 

In addition, the asymptotic flatness requirement imposes the following constraints on the mass function and Ricci scalar at infinity
\begin{align}\label{BC3}
    \lim_{r \rightarrow \infty} R(r) &= 0 ,   &   \lim_{r \rightarrow \infty} m(r) = \rm constant ,
\end{align}
Finally, the central value of the metric variable $\psi$ in Eq.~(\ref{BC1}) is fixed by requiring that the spacetime geometry be asymptotically flat, namely $\psi(r\rightarrow \infty) \rightarrow 0$. Thus, by taking into account Eq.~(\ref{Mf2}), the total gravitational mass of the star $M$ will be calculated from the asymptotic behavior
\begin{equation}\label{28}
    M \equiv \lim_{r \rightarrow \infty} \frac{r}{2}\left( 1- \frac{1}{e^{2\lambda}} + \frac{q^2}{r^2} \right) .
\end{equation}

\subsection{Starobinsky model }

The stellar structure equations (\ref{TOV1})-(\ref{TOV5}) require specifying a particular form for the Lagrangian function. Although mathematically any function could be considered, it is clear that there must be a physical motivation that justifies a particular choice. It is well known that the $2018$ release of data from the \textit{Planck} mission currently provide the best constraints on the CMB anisotropies \cite{Akrami2020}. In that regard, here we will consider the non-linear function $f(R)= R+ \alpha R^2$, whose inflationary predictions are in good agreement with \textit{Planck} $2018$ data. It is also known as the Starobinsky model \cite{Starobinsky1980} where $\alpha$ is the only free parameter of the theory and is usually given in units of $r_g^2$, where $r_g = GM_\odot/c^2 \approx 1.477\ \text{km}$ is the solar mass in geometrical units. At astrophysical scale, such a model has been widely used to investigate compact stars by means of perturbative \cite{Cooney2010, Arapoglu2011, Orellana2013, Astashenok2013, Resco2016} and non-perturbative \cite{Yazadjiev2014, Astashenok2015, Yazadjiev2015, Astashenok2017, Sbisa2020, Astashenok2020, Astashenok2021, Jimenez2021} approaches. See also Ref.~\cite{Nobleson2022} for a comparison of results on the mass-radius diagrams of neutron stars predicted by both methods. Here, for the first time, we address the study of compact stars with electric charge in $f(R)$ gravity theories using a non-perturbative method for the field equations.

Let us now see why the value of $\alpha$ has to be positive. In the exterior region of the star, where both energy density and pressure vanish, Eq.~(\ref{TOV3}) yields
\begin{equation}\label{RicciEq}
    R''+ \left( \psi'- \lambda' + \frac{2}{r} \right)R' - \frac{e^{2\lambda}}{6\alpha}R = 0 .
\end{equation}
It is to be expected that the adopted model coincides with the GR solution at sufficiently large distances from the stellar surface where $R(r) \rightarrow 0$. In addition, since the metric functions $\psi$ and $\lambda$ (as well as their first derivatives) asymptotically go to zero, we can approximate Eq.~(\ref{RicciEq}) as follows
\begin{equation}
    R'' + \frac{2}{r}R' - \frac{R}{6\alpha} = 0 ,
\end{equation}
whose solution is given by
\begin{equation}
    R(r)= \frac{a_1}{r}e^{-r/\sqrt{6\alpha}} + \frac{a_2}{r}\sqrt{\frac{3\alpha}{2}}e^{r/\sqrt{6\alpha}} ,
\end{equation}
where $a_1$ and $a_2$ are integration constants. We set $a_2= 0$ because the Ricci scalar must go to zero at infinity. Thus, in the vacuum, the function $R(r)$ decays approximately as
\begin{equation}
    R(r)= \frac{a_1}{r}e^{-r/\sqrt{6\alpha}} .
\end{equation}

It can now be seen that negative values of $\alpha$ lead to an oscillatory solution and which does not represent a physical behavior for the Ricci scalar. Therefore, we will only consider positive values in our analysis. 

It is worth mentioning that the inflaton mass fixed by CMB observations is given by $M \sim 10^{-5}\, M_{\rm Pl}$ or $M \sim 10^{13}\, \rm GeV$ \cite{Ivanov2022, Ketov2022}. Taking this into account, we have $\alpha = 1/(6M^2) \sim 10^{-60}\, \rm m^2$. In the present work, we will use values for $\alpha$ of the order of $r_g^2 \sim 10^{6}\, \rm m^2$, since are typical values used in the literature to construct compact stars in the Starobinsky model \cite{Yazadjiev2015, Astashenok2015, Astashenok2017, Jimenez2021, Nobleson2022}. This means that the value fixed by the CMB measurements is much less than our value used in the generation of charged quark stars. As can be observed from the mass-radius diagrams (see Figs.~\ref{figure4} and \ref{figure5}), the value fixed by the CMB would not generate appreciable changes (compared to the pure GR solutions) on the basic properties of a compact star.

\section{Equation of state and charge model }\label{Sec3}

As already mentioned, solving the modified TOV equations requires adopting an equation of state, i.e., a specific relation between energy density and pressure. In order to explore electrically charged quark stars, here we employ the MIT bag model EoS, which is given by 
\begin{equation}\label{MITEoS}
   p = \dfrac{1}{3}\left(\rho - 4B\right),
\end{equation}
where the parameter $B$ is the bag constant and it lies in the range $57 \leq B \leq 92\, \rm MeV/fm^3$. In this work, we will use $B = 57\, \rm MeV/fm^3$. Such value is commonly used because it generates maximum-mass configurations of about $2M_\odot$ within the context of Einstein gravity. We keep the same value for comparison of our results with those from the general relativistic case.

We are conscious of the sensitivity of structural characteristics of the compact stellar objects to different values of the MIT bag constant in the quark-matter EoS. However, this discussion is not the focus of the comparative study present here. We note that recently \cite{Delgado2022}, values for $B$ are consistently determined together with other parameters of the proposed model solving the TOV equations. The authors obtained the total charge $Q$ of compact stars, considering the good knowledge of mass and radius of a dozen selected objects. Thus the MIT bag constant, the central density,  surface density and central pressure are extracted for each object, presenting acceptable values. They are taken as structural model parameters for the specified objects. Nevertheless, in the present paper, we focus on the dependence of  physical quantities (mass, radius) on $\alpha$ and $\beta$ rather than on $B$.

In addition, we need to specify a charge density profile. Following the discussion in Refs. \cite{Ray2003, Arbanil2013}, we assume that the electric charge density is proportional to the energy density of standard matter, namely
\begin{equation}\label{CDprofile}
  \rho_{\rm ch} = \beta \rho ,
\end{equation}
where the constant $\beta$ is a charge parameter and it controls the amount of charge within the charged fluid.
We point out that a surface charge (which would be expected in a conductive material) would generate no Electric field in the bulk and, hence, there would be no change in the structure of the star itself.


\section{Numerical results}\label{Sec4}

Using the specific function $f(R)= R+ \alpha R^2$, equation of state (\ref{MITEoS}) and charge profile (\ref{CDprofile}), we numerically solve the modified TOV equations (\ref{TOV1})-(\ref{TOV5}) with the boundary conditions (\ref{BC1})-(\ref{BC3}). Namely, for the interior problem we integrate the differential equations from the origin up to the surface of the star where the pressure vanishes, and for the exterior problem we integrate from the surface up to infinity\footnote{Actually, we integrate up to a certain value of the radial coordinate from which the mass becomes constant (up to $0.01\%$).} fulfilling both the junction conditions (\ref{BC2}) and the asymptotic flatness requirement (\ref{BC3}). This can be done as long as the Ricci scalar assumes a unique value at the stellar center. 

\begin{figure*}
  \includegraphics[width=0.485\textwidth]{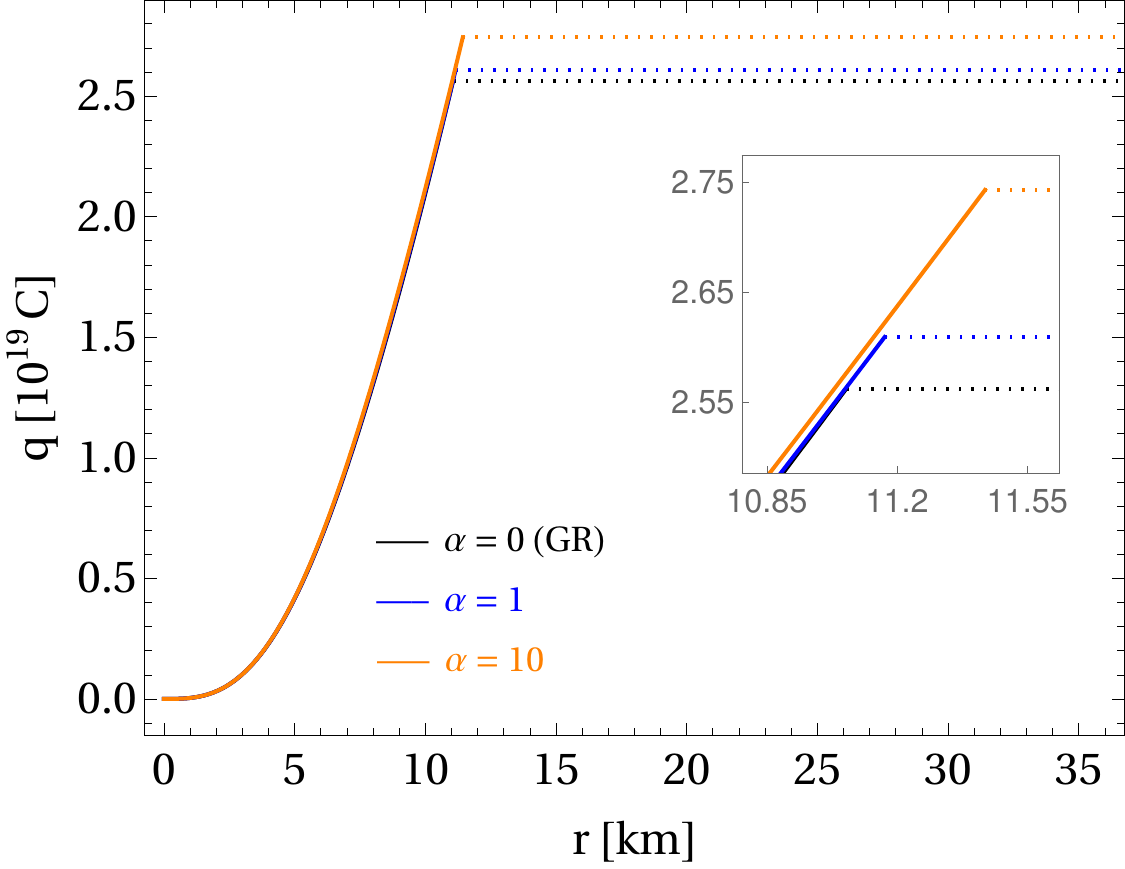}
  \includegraphics[width=0.485\textwidth]{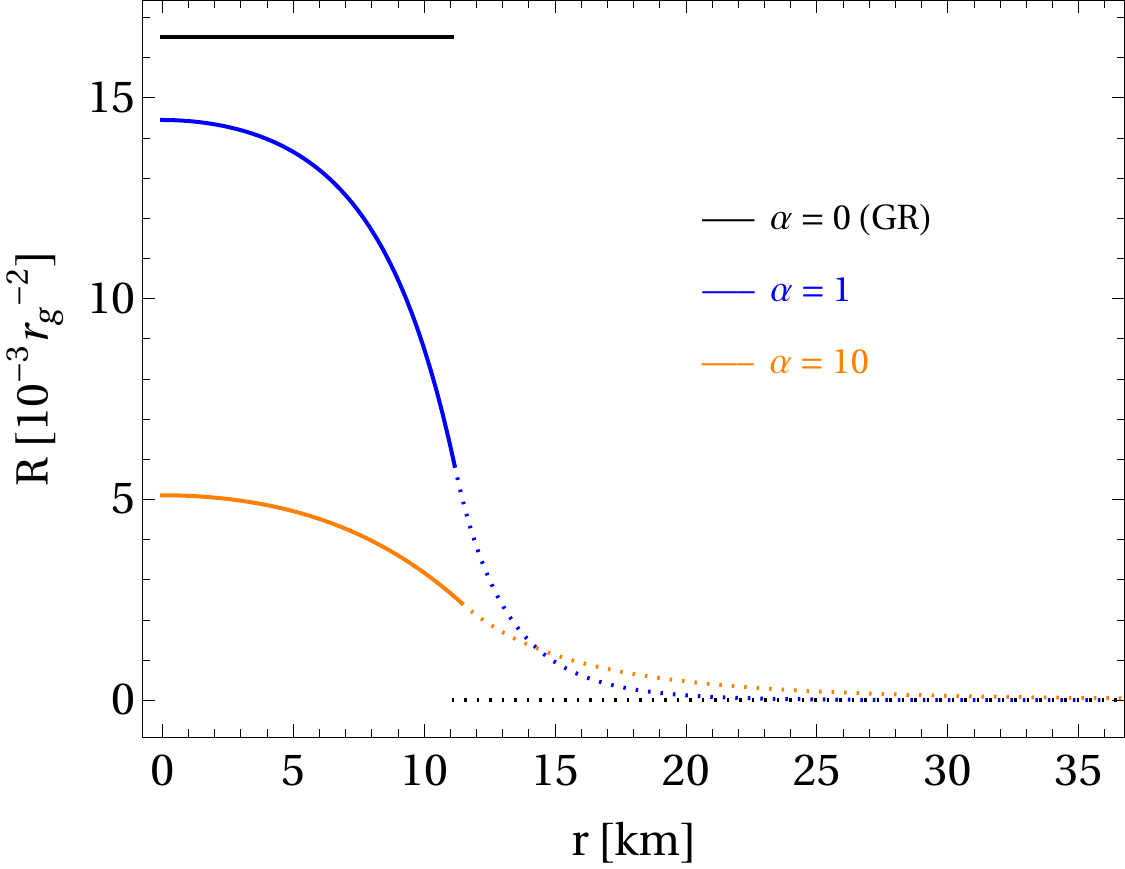}
  \includegraphics[width=0.485\textwidth]{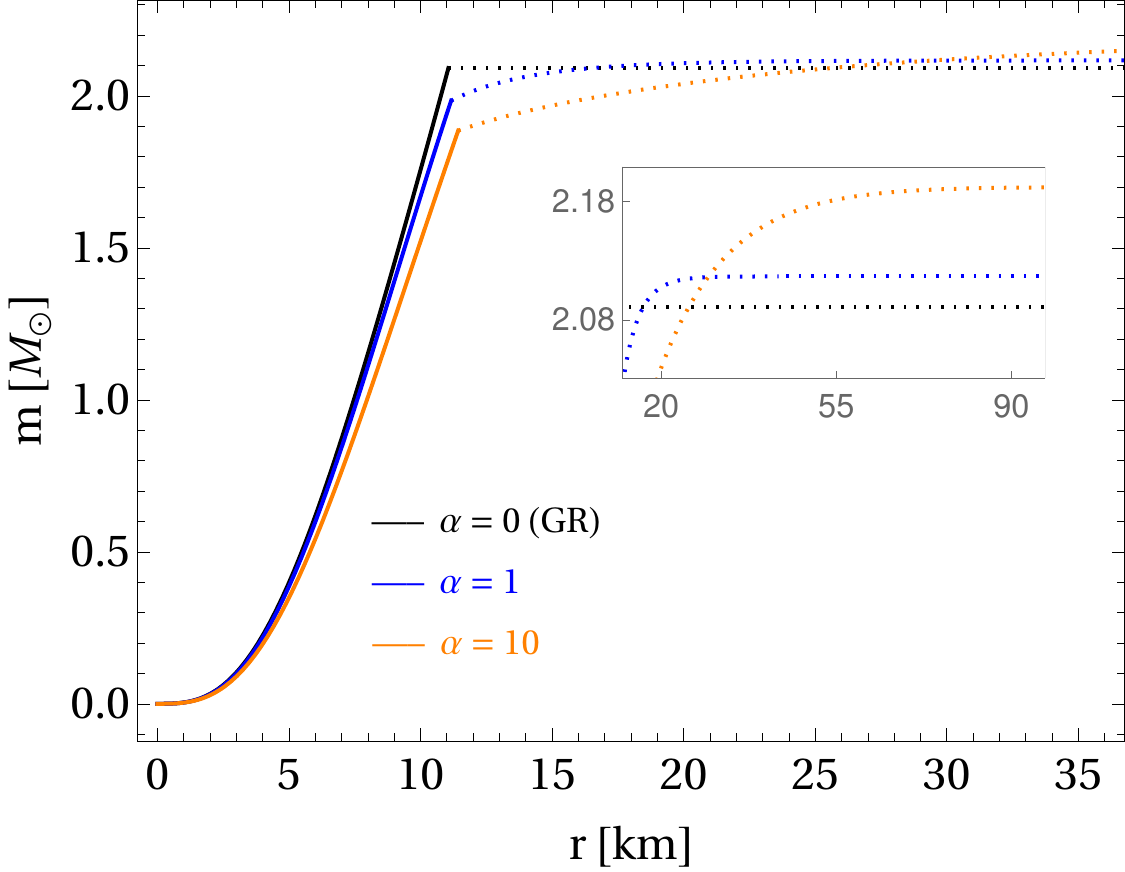}
  \includegraphics[width=0.486\textwidth]{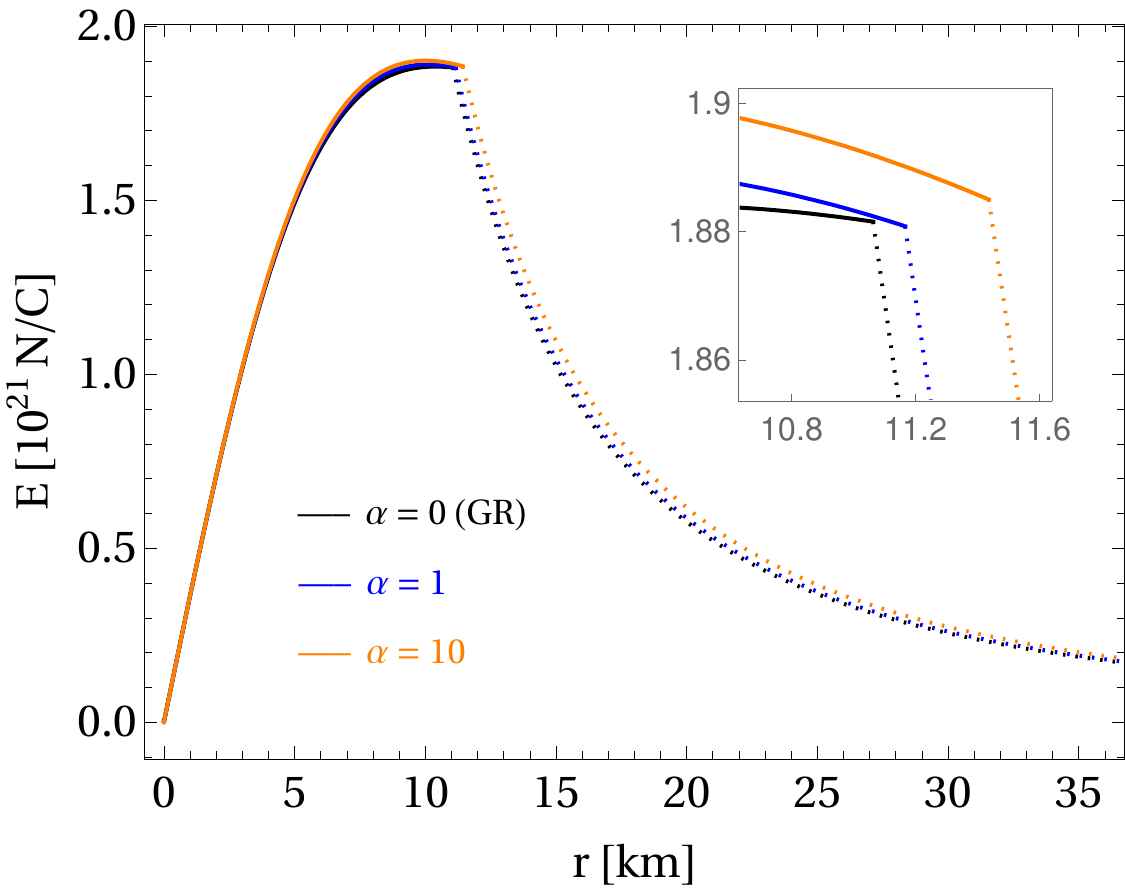}
  \caption{\label{figure1} Numerical solution of the modified TOV equations (\ref{TOV1})-(\ref{TOV5}) for the MIT bag model EoS (\ref{MITEoS}) and central energy density $\rho_c = 2.0 \times 10^{18}\, \rm kg/m^3$. The graphs are plotted for $\beta = 0.2$ and different values of the free parameter $\alpha$. The black curve represents the pure general relativistic solution. The solid and dotted lines correspond to the interior and exterior solutions, respectively. {\bf \textit{Top left:}} Electric charge distribution inside and outside the star. {\bf \textit{Top right:}} Radial behavior of the Ricci scalar, where $R$ satisfies the asymptotic flatness requirement at infinity. {\bf\textit{Left bottom:}} Gravitational mass versus radial coordinate. It can be observed that, as $\alpha$ increases, the mass measured at the surface decreases but the mass at infinity increases with respect to the GR value. Note that the mass takes a constant value at infinity. {\bf \textit{Right bottom:}} Electric field plotted as a function of the radial coordinate. Notice that this quantity decreases as we move away from the star, as expected. }  
\label{4plots}    
\end{figure*}

For instance, for a given central energy density $\rho_c = 2.0 \times 10^{18}\, \rm kg/m^3$ and charge parameter $\beta= 0.2$, Fig.~\ref{figure1} displays the numerical solution of the stellar-structure differential equations for two values of $\alpha$ (including the pure GR case for comparison). The (total) charge grows as we approach the surface and is constant outside the star, as expected according to the boundary conditions (\ref{BC2}). It can be observed that the effect of the $\alpha R^2$ term is an increase in the radius and the total charge with respect to the value obtained in Einstein gravity. Although in GR the Ricci scalar $R$ is constant due to the EoS --- since $R = 8\pi(\rho- 3p)$ in GR, $R= 32\pi B = \rm cte.$ inside the star according to Eq.~(\ref{MITEoS})  ---  it is a decreasing function of $r$ in the Starobinsky model and, as a consequence of the asymptotic flatness requirement, it goes to zero as we move away from the surface. The mass measured at the surface decreases as $\alpha$ increases, however, the third integral of Eq.~(\ref{Mf3}) contributes an extra factor that makes the asymptotic mass increase with respect to its GR value. Finally, the radial behavior of the electric field, i.e. $E(r) = q(r)/4\pi\epsilon_0 r^2$, is shown in the bottom right panel of Fig.~\ref{figure1}.

In order to better analyze the effect of the higher-order terms on the stellar structure, we will describe the curvature-induced terms in the action as an effective fluid --- also known in the literature as ``curvature fluid''. This interpretation is possible if one writes the field equations (\ref{FielEq}) as $G_{\mu\nu} = 8\pi(\mathcal{M}_{\mu\nu} + \mathcal{E}_{\mu\nu} + T^{\rm eff}_{\mu\nu})$, where $G_{\mu\nu}$ is the usual Einstein tensor and $T^{\rm eff}_{\mu\nu}$ is the effective energy-momentum tensor, given by
\begin{align}
    T^{\rm eff}_{\mu\nu} \equiv&\ \dfrac{1}{8\pi}\left[ (1 - f_R)R_{\mu\nu} + \dfrac{1}{2}(f -R)g_{\mu\nu}  \right.  \nonumber \\
    &\hspace{0.8cm} + \nabla_\mu\nabla_\nu f_R - g_{\mu\nu}\square f_R \bigg] .
\end{align}
which can be put in the form of an energy-momentum tensor corresponding to an anisotropic perfect fluid, i.e., $T_\mu^{(\rm eff) \nu} = \text{diag}\left( -\rho^{\rm eff}, p^{\rm eff}, p^{\rm eff}_t, p^{\rm eff}_t \right)$, so that the curvature energy density takes the form 
\begin{align}\label{CurvEnerDensEq}
\rho^{\rm eff} =&\ \dfrac{1}{8\pi}\left\lbrace \dfrac{(1- f_R)}{r^2}\dfrac{d}{dr}\left[ r\left( 1- e^{-2\lambda} \right)\right] + \dfrac{1}{2}(Rf_R - f) \right.  \nonumber  \\
&\left. + \dfrac{1}{e^{2\lambda}}\left[ \left( \dfrac{2}{r}- \lambda' \right)f'_R + f''_R \right] \right\rbrace .
\end{align}

\begin{figure}
   \includegraphics[width=0.46\textwidth]{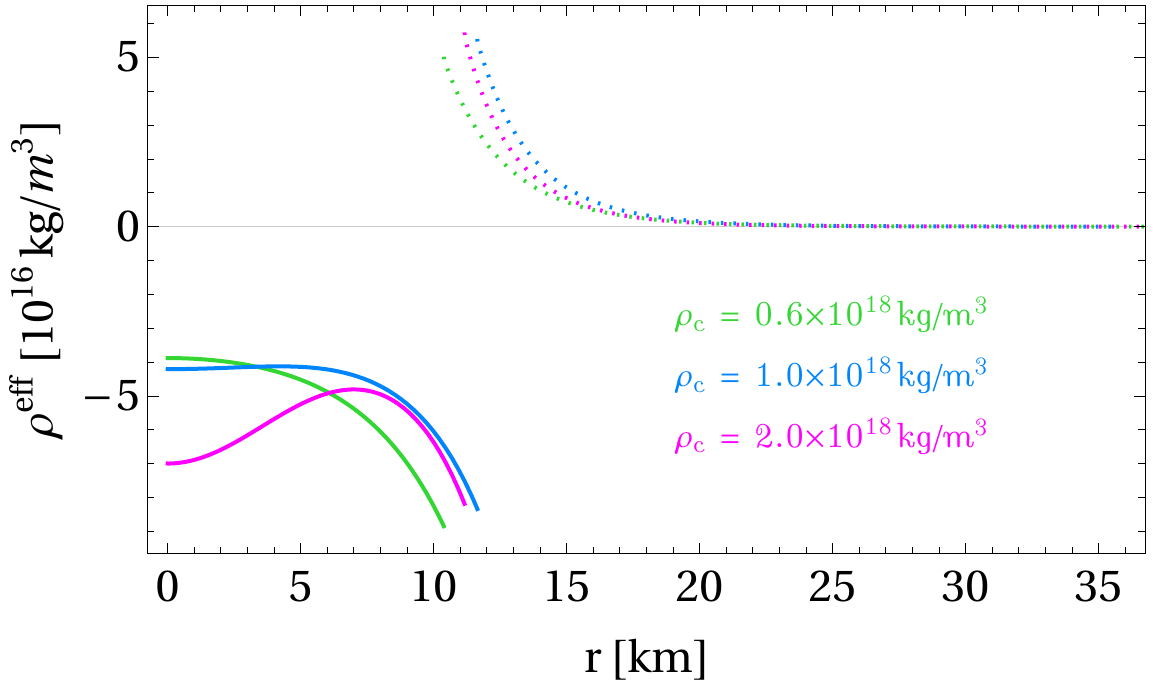}
   \caption{\label{figure2} Radial behavior of the effective energy density (\ref{CurvEnerDensEq})  for $\alpha= 1 r_g^2$, charge parameter $\beta= 0.2$ and three values of central energy density as indicated. A remarkable feature is that it is negative (positive) in the interior (exterior) of the charged star. }
\end{figure}

In view of Eq.~(\ref{CurvEnerDensEq}), we can define an effective mass as 
\begin{equation}\label{EffMassEq}
    m^{\rm eff}(r) = 4\pi\int_0^r \rho^{\rm eff}(\bar{r})\bar{r}^2d\bar{r} ,
\end{equation}
which corresponds precisely to the third integral given in Eq.~(\ref{Mf3}). The radial profile of the effective energy density for $\alpha= 1 r_g^2$ and $\beta= 0.2$ is exhibited in Fig.~\ref{figure2}. Its behavior is non-trivial, being negative (positive) inside (outside) the star regardless of the central energy density. According to Eq.~(\ref{EffMassEq}), such behavior leads to a negative effective mass in the interior of the star and, therefore, this explains why the total mass measured at the surface decreases as $\alpha$ increases, as shown in the bottom left panel of Fig.~\ref{figure1}. On the other hand, the effective mass is positive in the outer region, which makes the total mass at infinity exceed the GR counterpart for $\rho_c = 2.0 \times 10^{18}\, \rm kg/m^3$.

\begin{figure*}
   \includegraphics[width=0.47\textwidth]{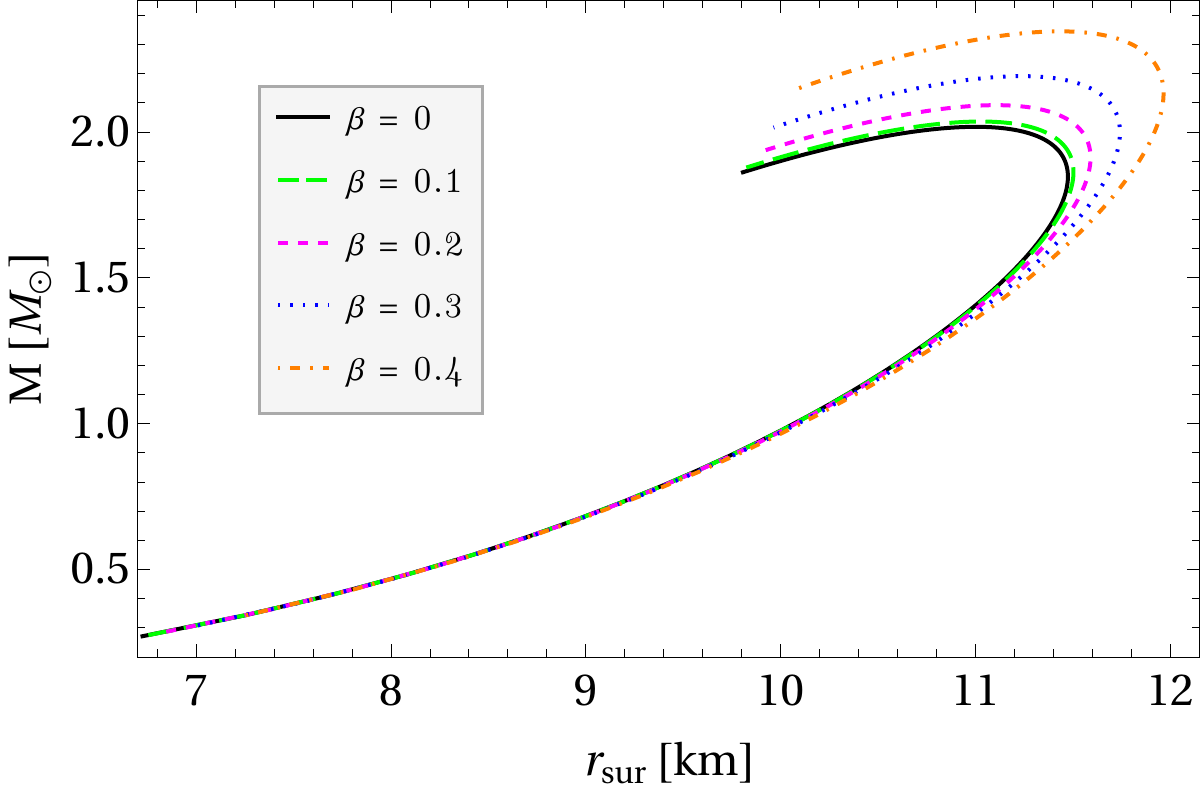}
   \includegraphics[width=0.46\textwidth]{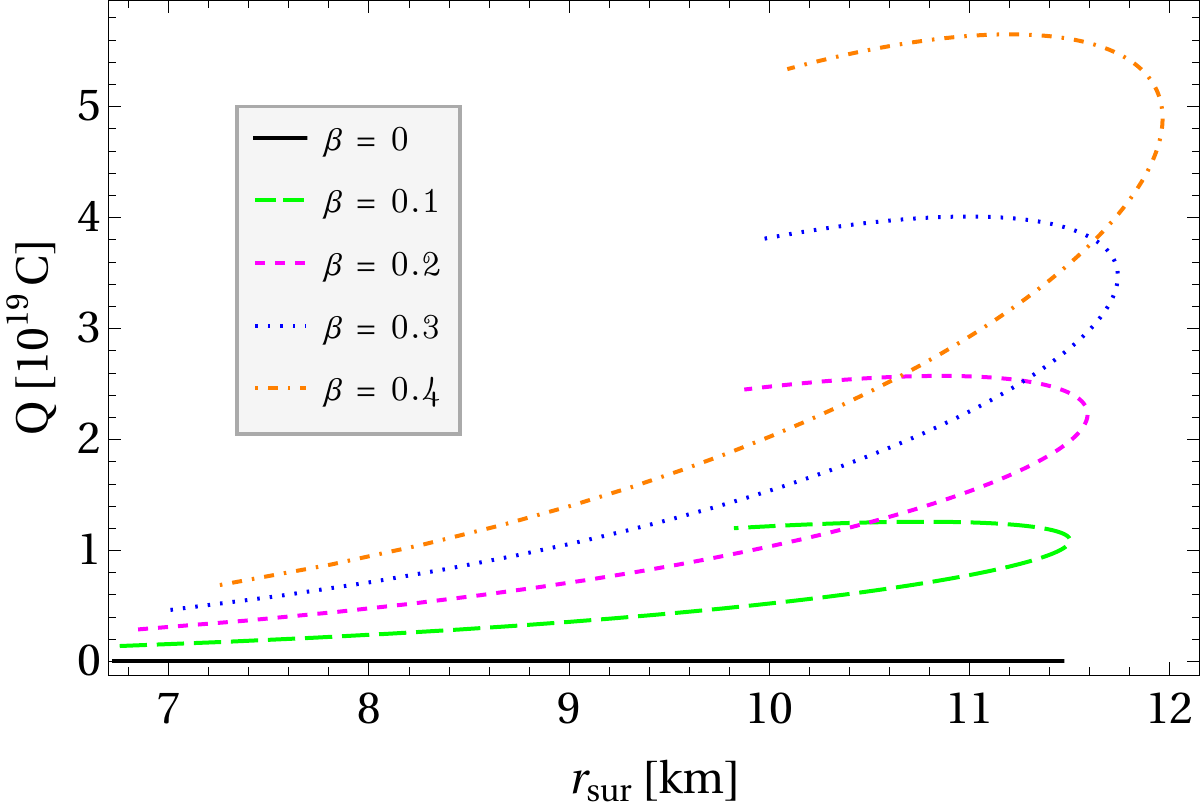}
   \caption{\label{figure3} Macro-physical properties of charged quark stars predicted by Einstein theory (i.e., when $\alpha =0$). Mass-radius diagram (left panel) and total charge versus radius (right panel) for different values of the charge parameter $\beta$. We see that the main consequence of the electrical charge contained in the stellar fluid is an increase in the maximum-mass values. Note that the most significant changes occur in the high-mass region. }
\end{figure*}

By varying the central energy density we can obtain a family of charged quark stars. In that regard, we have also plotted the (total) mass $M$ and (total) charge $Q$ as a function of the star radius $r_{\rm sur}$ in Fig.~\ref{figure3} for fixed $\alpha=0$ (i.e, GR) and varying $\beta$, in order to investigate  the effect of charge alone. As one can see, there is a significant (although actually unobservable by current standards) change in both $M$ and $r_{\rm sur}$. Obviously, $Q$ is strongly dependent on the charge parameter $\beta$.

\begin{figure*}
   \includegraphics[width=0.464\textwidth]{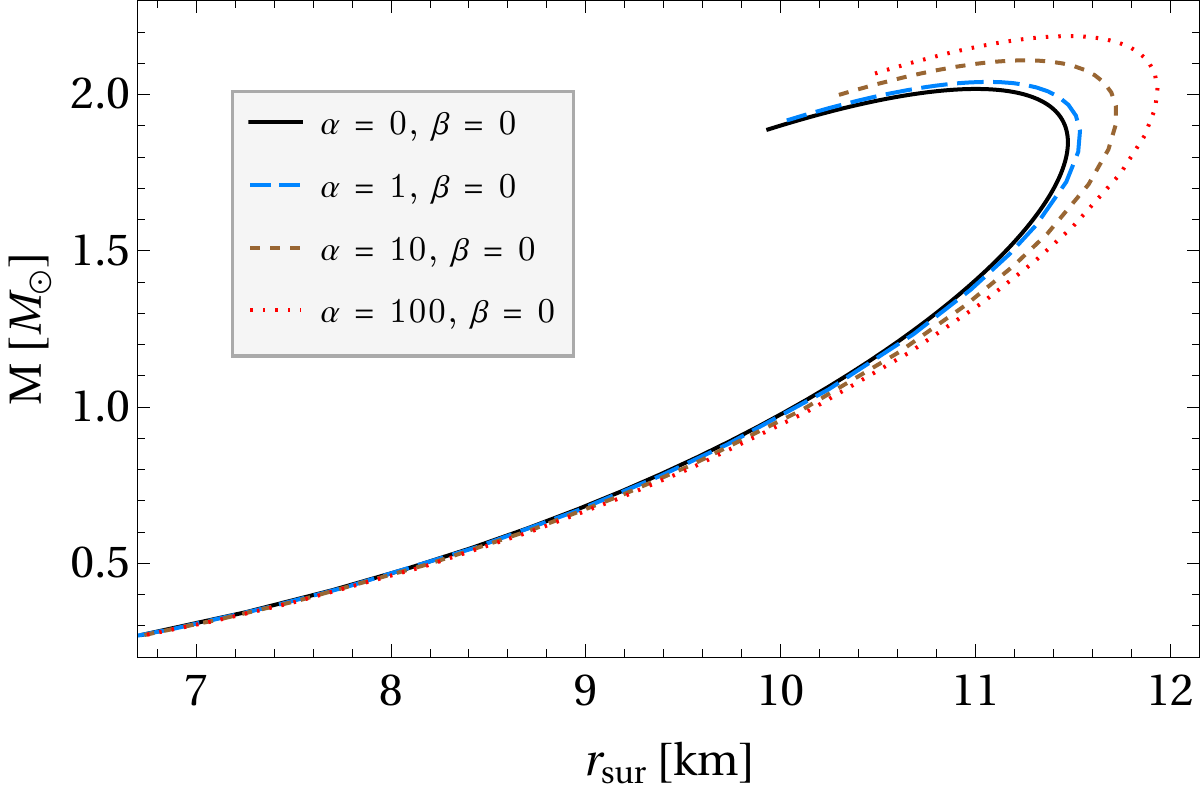}
   \includegraphics[width=0.46\textwidth]{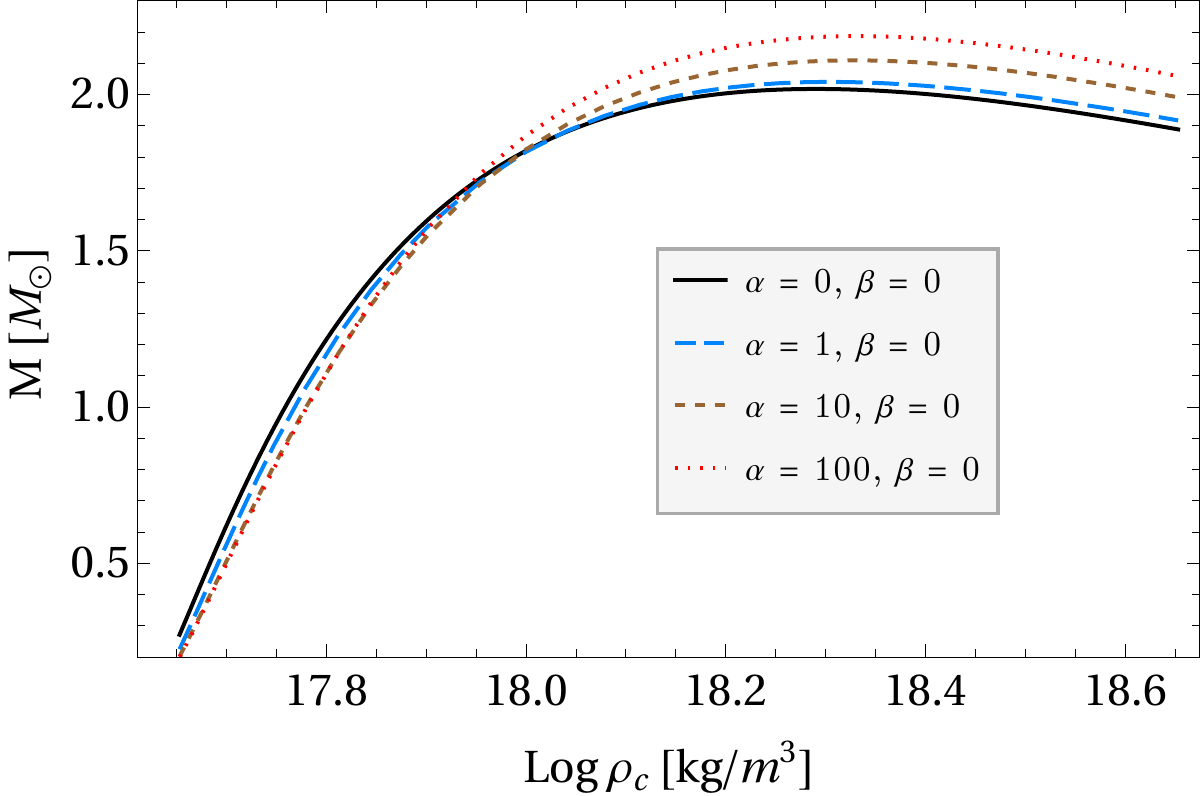}
   \caption{\label{figure4} Left panel: Mass versus radius relation as predicted by the Starobinsky model without electric charge (that is, when $\rho_{\rm ch} =0$ in the modified TOV equations), where four values of the free parameter $\alpha$ are particularly considered for the MIT bag model EoS (\ref{MITEoS}). Right panel: Gravitational mass as a function of the central energy density.  }
\end{figure*}

In Fig.~\ref{figure4}, we vary $\alpha$ but keep $\beta=0$, in order to point out the changes introduced by the $\alpha R^2$ term alone. The behaviour is qualitatively unaltered, with a slight increase in the maximum mass for larger values of the central density. We point out that our code reproduces previous results reported in the literature for uncharged quark stars with the so-called MIT bag model EoS within the Starobinsky model \cite{Astashenok2015}.

\begin{figure*}
   \includegraphics[width=0.464\textwidth]{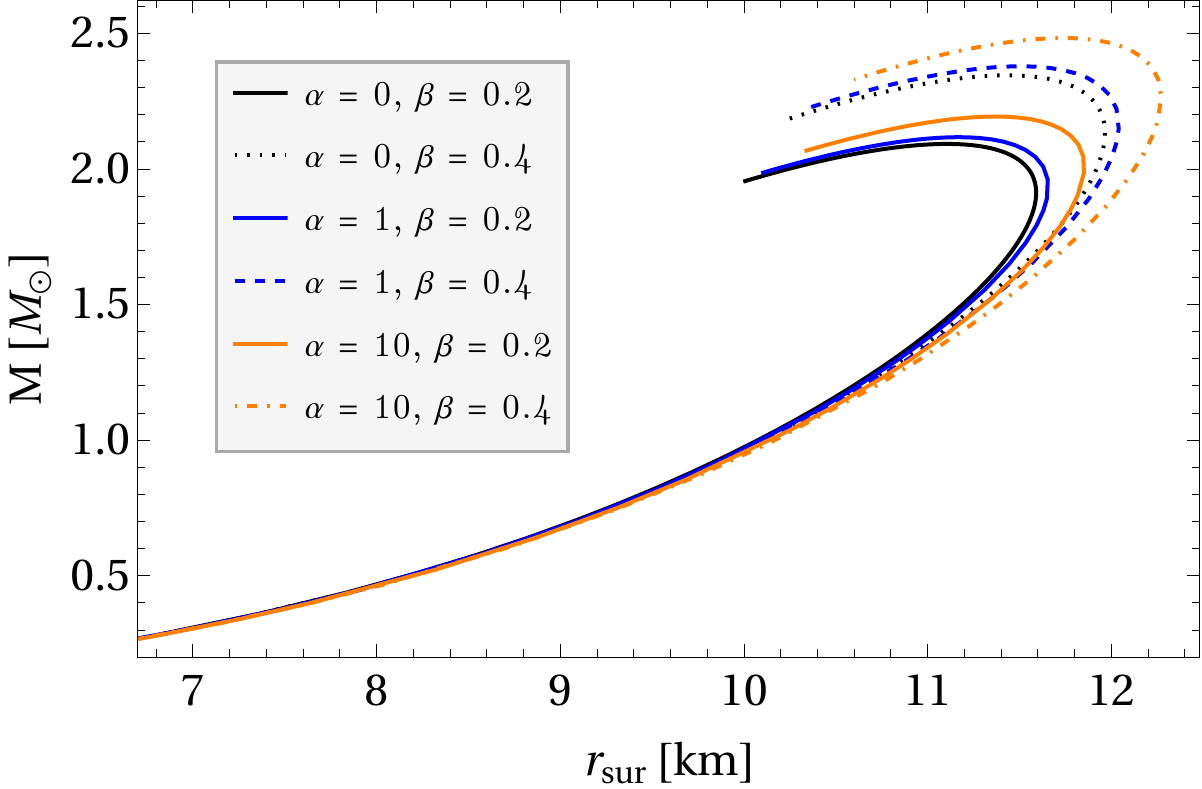}
   \includegraphics[width=0.46\textwidth]{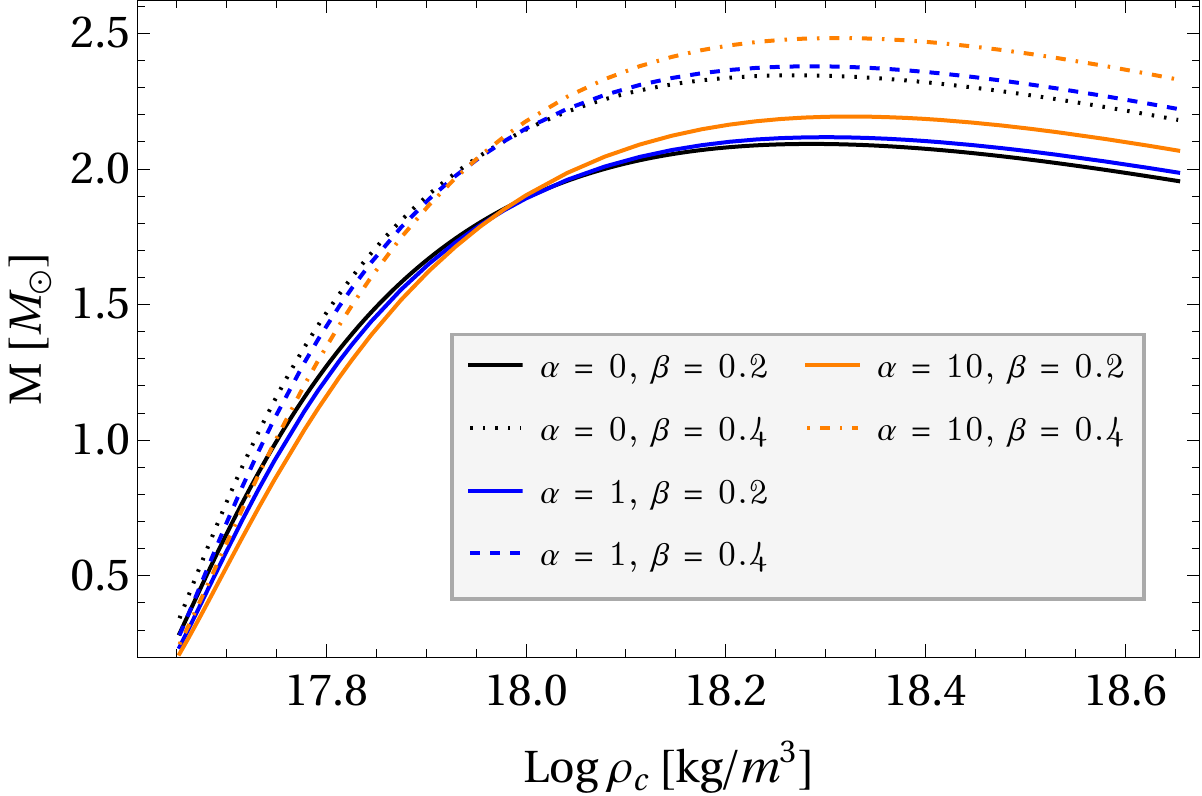}
   \caption{\label{figure5} Mass-radius diagrams (left panel) and mass-central density relations (right panel) for charged quark stars within the context of $f(R)= R+ \alpha R^2$ gravity. The black curves represent the pure GR case in both plots. In both panels, solid lines correspond to $\beta=0.2$, and dashed lines to $\beta=0.4$. In both sets, $\alpha$ varies from $0$ (GR) to $10$. }
\end{figure*}

\begin{figure}
   \includegraphics[width=0.46\textwidth]{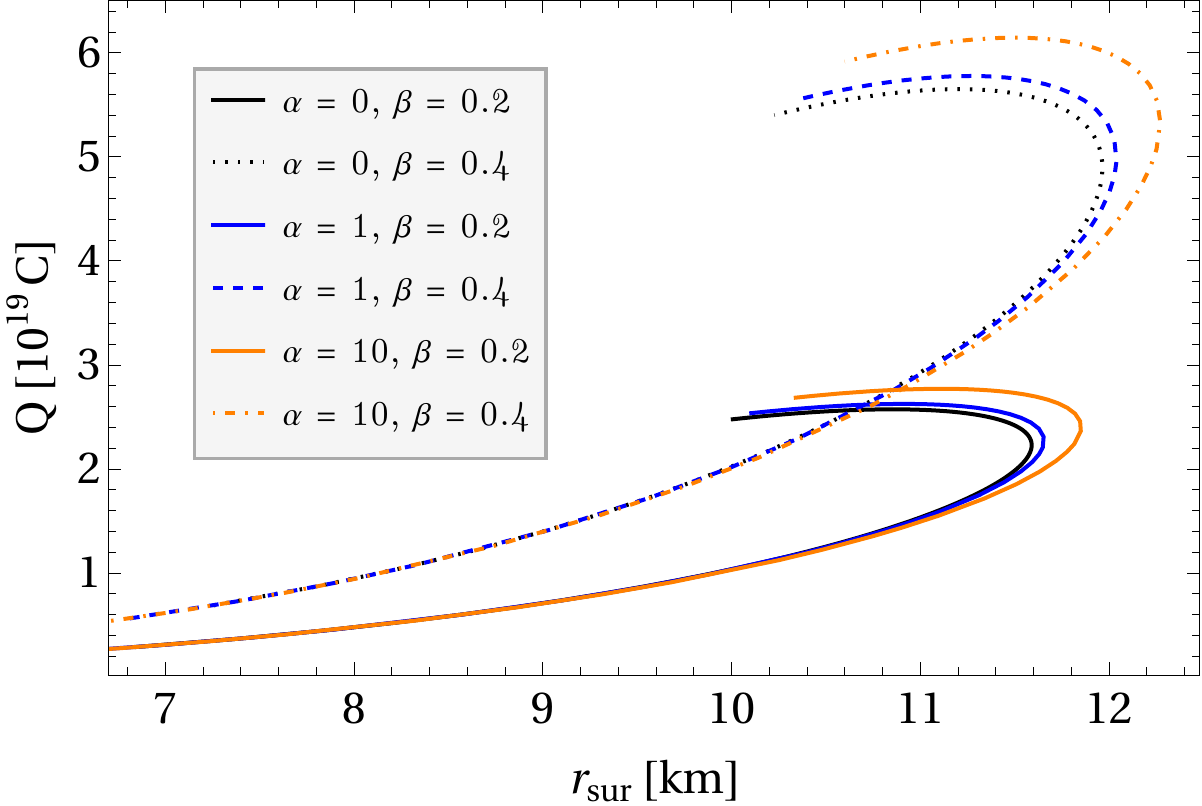}
   \caption{\label{figure6} Total charge as a function of the radius as predicted by the charge density profile (\ref{CDprofile}) in the Starobinsky model for different combinations of $\alpha$ and $\beta$. }
\end{figure}

Finally, we allow both $\alpha$ and $\beta$ to vary in Fig.~\ref{figure5} and \ref{figure6}. We can notice that the main parameter is indeed $\beta$, for it yields a larger change in both $M$ and $r_{\rm sur}$. In other words, both parameters allow an increase in the maximum-mass values, however, the greater effect is obtained by varying $\beta$. Furthermore, the most relevant changes in the total charge due to the quadratic term occur in the high-radius region.

\begin{figure*}
   \includegraphics[width=0.48\textwidth]{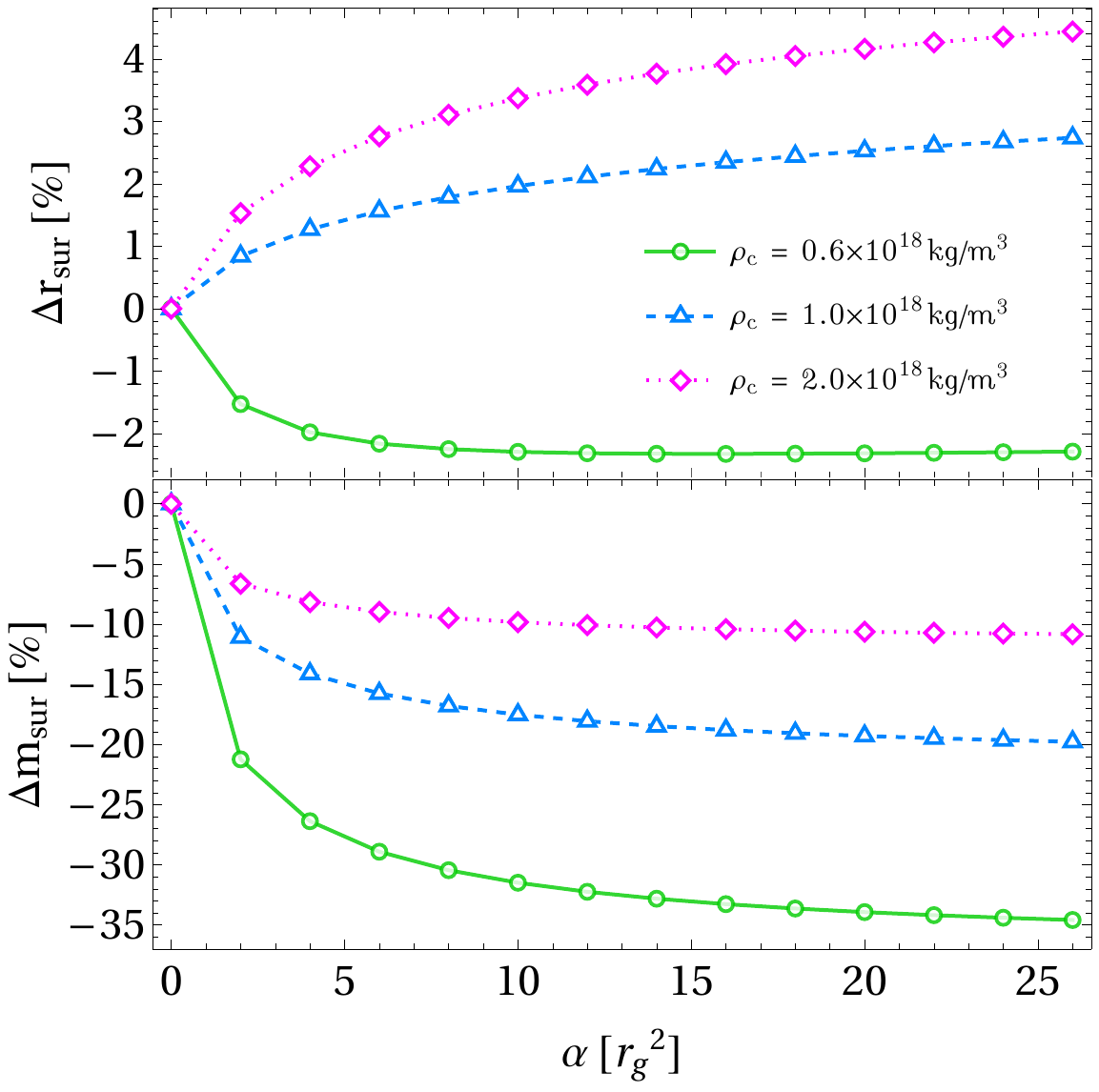}\,
   \includegraphics[width=0.48\textwidth]{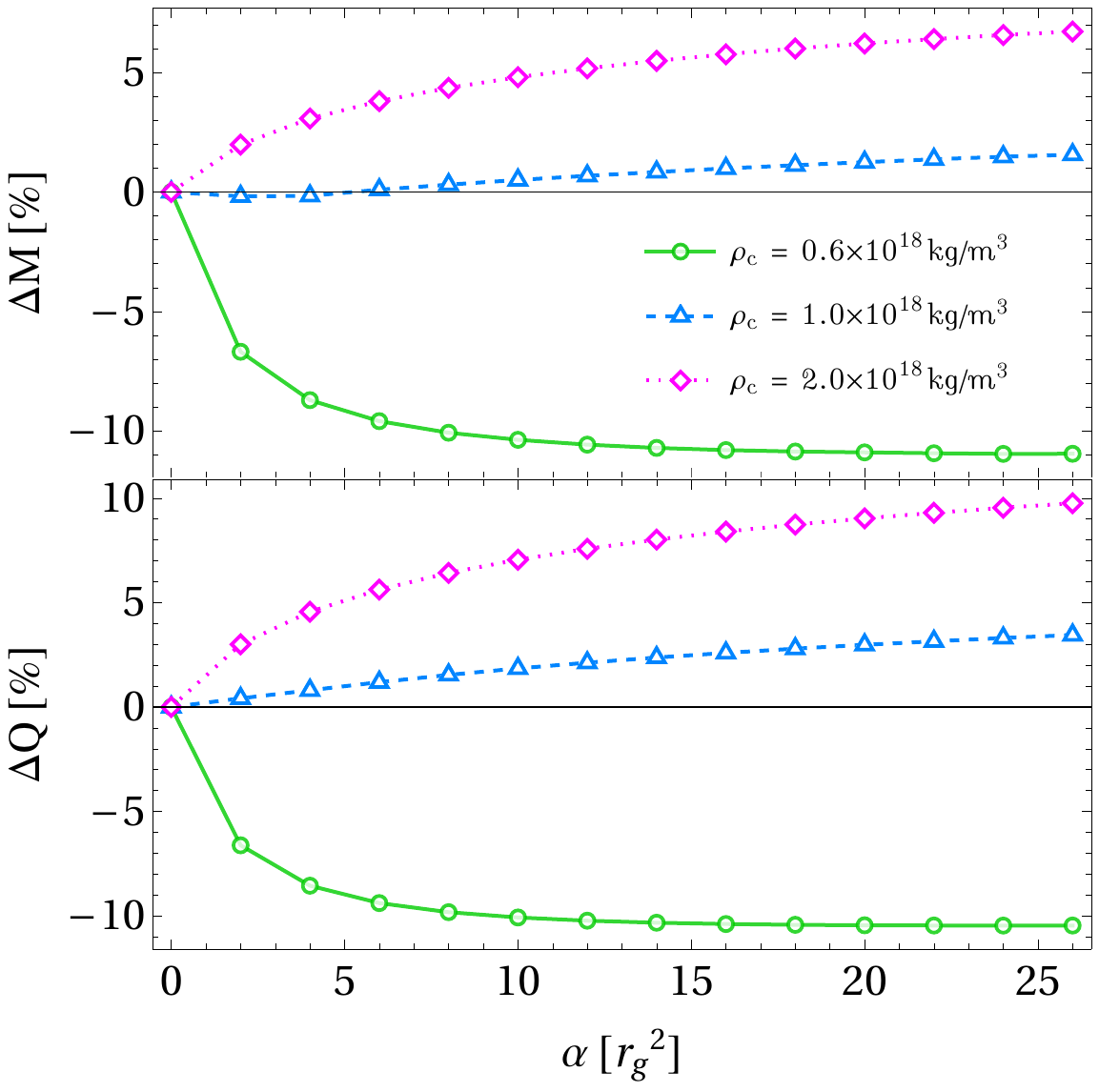}
   \caption{\label{RelativeFig} Relative deviation (\ref{RelativeEq}) as a function of the free parameter $\alpha$ for three values of central density and fixed $\beta=0.2$. For low central densities the radius decreases with respect to its value obtained in Einstein gravity, while the opposite occurs for higher central densities. The mass measured at the surface always decreases regardless of the central density value and variations can reach up to $35\%$. We also observe that the asymptotic mass and the total charge have a similar behavior, reaching deviations of up to $10\%$. }
\end{figure*}

In order to better quantify the deviations caused by the $\alpha R^2$ term on the macro-physical properties of a charged compact star with respect to the pure GR case, we define the following relative deviation for a given central density
\begin{equation}\label{RelativeEq}
    \Delta = \left. \frac{\Upsilon_{f(R)} - \Upsilon_{\rm GR}}{\Upsilon_{\rm GR}} \right\vert_{\rho_c} ,
\end{equation}
where $\Upsilon$ stands for any physical quantity. These deviations are shown in Fig.~\ref{RelativeFig} for three different values of central density and $\beta= 0.2$. The radius exhibits relative deviations of up to $4\%$ ($2\%$) for high (low) values of central energy density. The mass measured at the surface always decreases with respect to its GR counterpart and varies up to about $35\%$ for low densities. The discrepancies for the asymptotic mass can reach up to $10\%$ in the low-density region, although a less trivial deviation occurs for $\rho_c = 1.0 \times 10^{18}\, \rm kg/m^3$ where the mass can decrease and increase after a certain value of $\alpha$. Moreover, the total charge undergoes variations similar to the asymptotic mass.

\section{Concluding remarks}\label{Sec5}

In this work, we have investigated the effect of electric charge on the global properties of compact stars within the context of fourth-order $f(R)$ theories of gravity. Our goal was therefore to derive the modified TOV equations with the inclusion of the electromagnetic energy-momentum tensor under a non-perturbative approach to the field equations. We have considered the MIT bag model EoS for the dense matter involved and adopted a charge profile where the electric charge density is proportional to the standard energy density, giving rise to a parameter $\beta$ which controls the amount of charge inside the star.

For the $f(R)= R+ \alpha R^2$ gravity model, where $\alpha$ is a free parameter that measures the deviation from GR, we have numerically solved the stellar structure equations and hence we obtained the radius, surface mass, asymptotic mass and total charge for a large family of quark stars. We analyzed the impact of the $\alpha R^2$ term by means of an effective fluid, which led to a negative effective mass inside the star. As a consequence, the total mass measured at the surface decreases as $\alpha$ increases. Furthermore, we have introduced a relative deviation in order to better quantify the changes caused by the quadratic term with respect to the GR counterpart. We found that the radius increases (decreases) for high (low) central energy densities, while the mass measured at the surface always decreases with $\alpha$ increasing. The asymptotic mass and the total charge can decrease or increase depending on the central density value, suffering variations of up to $10\%$ for high values of $\alpha$. According to the total charge versus radius relation, the most substantial changes due to the Starobinsky term occur in the high-radius region. Nevertheless, it is worth emphasizing that the largest changes in radius, mass, and charge are due to the parameter $\beta$.

\begin{acknowledgments}
JMZP acknowledges financial support from the PCI program of the Brazilian agency ``Conselho Nacional de Desenvolvimento Científico e Tecnológico''--CNPq. JDVA thanks Universidad Privada del Norte and Universidad Nacional Mayor de San Marcos for the financial support - RR Nº$\,005753$-$2021$-R$/$UNMSM under the project number B$21131781$. SBD thanks CNPq for partial financial support. RRRR thanks CNPq for partial financial support (grant no. $309868/2021-1$). This work has been done as a part of the Project INCT-Física Nuclear e Aplicações, Project number $464898/2014-5$.  
\end{acknowledgments}\


%

\end{document}